\documentstyle{mn}

\newif\ifAMStwofonts
\ifoldfss
  \ifCUPmtlplainloaded \else
    \NewTextAlphabet{textbfit} {cmbxti10} {}
    \NewTextAlphabet{textbfss} {cmssbx10} {}
    \NewMathAlphabet{mathbfit} {cmbxti10} {} % for math mode
    \NewMathAlphabet{mathbfss} {cmssbx10} {} %  "   "    "
  \fi
  \ifAMStwofonts
    \ifCUPmtlplainloaded \else
      \NewSymbolFont{upmath} {eurm10}
      \NewSymbolFont{AMSa} {msam10}
      \NewMathSymbol{\upi}     {0}{upmath}{19}
      \NewMathSymbol{\umu}     {0}{upmath}{16}
      \NewMathSymbol{\upartial}{0}{upmath}{40}
      \NewMathSymbol{\leqslant}{3}{AMSa}{36}
      \NewMathSymbol{\geqslant}{3}{AMSa}{3E}

    \fi
  \fi
\fi % End of OFSS

\ifnfssone
  \newmathalphabet{\mathit}
  \addtoversion{normal}{\mathit}{cmr}{m}{it}
  \addtoversion{bold}{\mathit}{cmr}{bx}{it}
  \newmathalphabet{\mathbfit} % math mode version of \textbfit{..}
  \addtoversion{normal}{\mathbfit}{cmr}{bx}{it}
  \addtoversion{bold}{\mathbfit}{cmr}{bx}{it}
  \newmathalphabet{\mathbfss} % math mode version of \textbfss{..}
  \addtoversion{normal}{\mathbfss}{cmss}{bx}{n}
  \addtoversion{bold}{\mathbfss}{cmss}{bx}{n}
  \ifAMStwofonts
    \ifCUPmtlplainloaded \else
      \UseAMStwoboldmath
      \makeatletter
      \new@mathgroup\upmath@group
      \define@mathgroup\mv@normal\upmath@group{eur}{m}{n}
      \define@mathgroup\mv@bold\upmath@group{eur}{b}{n}
      \edef\UPM{\hexnumber\upmath@group}
      \new@mathgroup\amsa@group
      \define@mathgroup\mv@normal\amsa@group{msa}{m}{n}
      \define@mathgroup\mv@bold\amsa@group{msa}{m}{n}
      \edef\AMSa{\hexnumber\amsa@group}
      \makeatother
      \mathchardef\upi="0\UPM19
      \mathchardef\umu="0\UPM16
      \mathchardef\upartial="0\UPM40
      \mathchardef\leqslant="3\AMSa36
      \mathchardef\geqslant="3\AMSa3E
    \fi
  \fi
\fi % End of NFSS release 1

\ifnfsstwo
  \DeclareMathAlphabet{\mathbfit}{OT1}{cmr}{bx}{it}
  \SetMathAlphabet\mathbfit{bold}{OT1}{cmr}{bx}{it}
  \DeclareMathAlphabet{\mathbfss}{OT1}{cmss}{bx}{n}
  \SetMathAlphabet\mathbfss{bold}{OT1}{cmss}{bx}{n}
  \ifAMStwofonts
    \ifCUPmtlplainloaded \else
      \DeclareSymbolFont{UPM}{U}{eur}{m}{n}
      \SetSymbolFont{UPM}{bold}{U}{eur}{b}{n}
      \DeclareSymbolFont{AMSa}{U}{msa}{m}{n}
      \DeclareMathSymbol{\upi}{0}{UPM}{"19}
      \DeclareMathSymbol{\umu}{0}{UPM}{"16}
      \DeclareMathSymbol{\upartial}{0}{UPM}{"40}
      \DeclareMathSymbol{\leqslant}{3}{AMSa}{"36}
      \DeclareMathSymbol{\geqslant}{3}{AMSa}{"3E}
    \fi
  \fi
\fi % End of NFSS release 2

\ifCUPmtlplainloaded \else
  \ifAMStwofonts \else % If no AMS fonts
    \def\upi{\pi}
    \def\umu{\mu}
    \def\upartial{\partial}
  \fi
\fi

\title{A spatially resolved `inside-out' outburst of IP Pegasi}
\author[Webb et\, al.]
       {N.A. Webb, $^{1}$\thanks{email:naw@astro.keele.ac.uk}  T. Naylor,$^{1}$  
 Z. Ioannou, $^{1}$  W.J. Worraker,$^{2}$ J. Stull, $^{3}$ A. Allan, $^{1}$ \cr
R. Fried,$^{4}$ N.D. James$^{5}$ and D. Strange $^{6}$ \\
$^{1}$Department of Physics, Keele University, Keele, Staffordshire, ST5 5BG, UK \\
$^{2}$65, Wantage Road, Didcot, Oxfordshire, OX11 0AE, UK \\
$^{3}$Stull Observatory, Alfred University, Alfred, NY 14802, USA \\
$^{4}$Braeside Observatory, P.O. Box 906 Flagstaff, Arizona 86002, USA \\
$^{5}$11, Tavistock Road, Chelmsford, Essex, CM1 6JL, UK \\
$^{6}$Worth Hill Observatory, Worth Matravers, Dorset UK}
\date{Accepted ???. 
      Received ???; 
      in original form ???}

\begin{document}

\maketitle

\label{firstpage}

\begin{abstract}
We present a comprehensive photometric dataset taken over the entire
outburst of the eclipsing dwarf nova IP Peg in September/October 1997.
Analysis of the lightcurves taken over the long rise to the
peak-of-outburst shows conclusively that the outburst started near the
centre of the disc and moved outwards.  This is the first dataset
that spatially resolves such an outburst. The dataset is consistent
with the idea that long rise times are indicative of such `inside-out'
outbursts.  We show how the thickness and the radius of the disc, along
with the mass transfer rate change over the whole outburst.  In
addition, we show evidence of the secondary and the irradiation
thereof.  We discuss the possibility of spiral shocks in the disc;
however we find no conclusive evidence of their existence in this dataset.
\end{abstract}


\section{Introduction}

IP Pegasi is a U-Gem type dwarf nova.  It has an orbital period of 3.8
hours and an average outburst cycle length of about 95 days.  Being
the brightest deeply eclipsing dwarf nova known with an orbital period
above the period gap, it is an essential object to study, to
understand the differences between long- and short-period systems.

From observations presented in this paper, the visible disc observed
in IP Peg at the onset of the September-October 1997
outburst was small and grew over a couple of days (see Sections
\ref{sec:phot} \& \ref{sec:modelec}).  This suggests that the
peak-of-outburst occurs when the heating wave reaches its maximum
distance from the white dwarf.  A cooling wave then moves inwards at a
slower rate, causing the visible part of the disc to decrease in
radius and thereby the system declines from outburst. Thus we observed
an inside-out outburst as described by Smak \shortcite{Smak84}, Ludwig
\& Meyer \shortcite{Ludw98} etc.

Recent work on IP Peg by Steeghs, Harlaftis \& Horne
\shortcite{Stee97} and Wolf et al \shortcite{Wolf98}, has shown the
possibility that IP Peg contains tidal shocks (or spiral arms) within
the disc.  Their presence has been determined by Doppler tomography
(Marsh \& Horne 1988).  Various lines in the visible spectra, such as
H$\alpha$ and He I ($\lambda$6678) are mapped from time-resolved
spectra, from which the velocity field is deduced and displayed on
maps where the primary's velocity coordinates are defined to be the
origin.  Such shocks have been discussed theoretically by e.g.
Sawada, Matsuda \& Hachisu \shortcite{Sawa86}, Savonije, Papaloizou \&
Lin \shortcite{Savo94}, Spruit et al \shortcite{Spru87} and Dgani,
Livio \& Regev \shortcite{Dgan94}.  They are tidally induced by the
secondary star and should therefore be evident as the disc approaches
the tidal radius, which IP Peg can be seen to do during this outburst.
The perturbations become coherent spiral shocks, which extend from the
outer disc down to quite small radii, depending on the Mach number of
the disc flow.  These arms are hypothesised to provide the removal of
angular momentum required to allow material to be able to fall onto
the white dwarf.  This then disposes with the need for a disc
viscosity, prescribed in the $\alpha$ Disc Instability Model, e.g
Meyer \& Meyer-Hofmeister \shortcite{Meye82}, Faulkner, Lin \&
Papaloizou \shortcite{Faul83} etc.  We find no unequivocal evidence
that spiral shocks existed in the disc during this outburst,
though we discuss this further in Section~\ref{sec:sparms}.

\vspace*{-0.5cm}

\section{Observations and data reduction}

The September - October 1997 outburst of IP Peg was observed in the V
band at several sites.  The observations are outlined in Table
\ref{tab:obs} and details of most of the instruments used are given in
Ioannou et al (1999), henceforth Paper~I.  The other instruments used
were a 0.4m classical Cassegrain reflector with a CCD camera, at
Braeside Observatory, Arizona, and a 0.81m, f/4 Newtonian reflector,
fitted with a ST-6B CCD camera, at Stull Observatory, New York State
(N.\ Y.\ S.).  We also observed the 1994 December outburst with UKIRT
and IRCAM-3.

\begin{table}
 \begin{center}
 \caption{V-band outburst observations of IP Peg Sep-Oct 1997}
 \begin{tabular}{|cccccc|}
 \hline
 Date & Ecl- & BDJD & Site & Int. & Cycle \\
 (1997) & ipse & (2450710+) & & time(s) & time(s)\\
 \hline
 Sep 21 & 4519 & 3.451259 & Keele & 10 & 13 \\
        & 4520 & 3.609384 & Keele & 10 & 13 \\
 Sep 22 & 4525 & 4.400453 & Keele & 3 & 7 \\
        & 4525 & 4.400453 & Essex & 15 & 20 \\
        & 4525 & 4.400453 & Dorset& 20 & 122 \\
        & 4526 & 4.558647 & Keele & 3 & 7 \\
 Sep 23 & 4531 & 5.349651 & Keele & 3 & 6 \\
 Sep 24 & 4538 & 6.457649 & Dorset& 20 & 122 \\
 Sep 27 & 4557 & 9.462988 & Keele & 3  & 6 \\
 Sep 28 & 4558 & 9.620064 & N.\ Y.\ S. & 10 & 13 \\
        & 4559 & 9.779402 & N.\ Y.\ S. & 10 & 13 \\
 Sep 30 & 4571 & 11.51671 & N.\ Y.\ S. & 10 & 13 \\
 Oct 01 & 4582 & 13.41814 & Keele & 3  & 6\\
 Oct 01 & 4582 & 13.41814 & Essex & 15 & 20 \\
 Oct 02 & 4585 & 13.73463 & N.\ Y.\ S. & 10 & 13 \\
 Oct 02 & 4588 & 14.36743 & Keele & 3 & 5 \\
 Oct 05 & 4607 & 17.37332 & Keele & 3  & 5 \\
 Oct 06 & 4608 & 17.53160 & N.\ Y.\ S. & 10 & 13 \\
        & 4609 & 17.68989 & N.\ Y.\ S. & 10 & 13 \\
        & 4609 & 17.68989 & Braeside & 60 & 71 \\
        & 4610 & 17.84792 & N.\ Y.\ S. & 10 & 13 \\
        & 4610 & 17.84792 & Braeside & 60 & 71 \\
 Oct 07 & 4616 & 18.63793 & Braeside & 60 & 71 \\
        & 4617 & 18.80035 & Braeside & 60 & 71 \\
 Oct 25$^a$ & 4734 & 37.46551 & Keele & 10 & 13 \\
 
 \hline
 \end{tabular}
 \end{center}
 \label{tab:obs}
 $^a$ quiescent observation
\end{table}

The optical photometry was reduced as described in Paper~I, the local
standard stars being those of Misselt (1996).  We checked for
differences in the colour response of the V band filters between
different telescopes by comparing simultaneous eclipse data, since the
star's colour changes most dramatically through eclipse.  Deviations
were within a few percent for the Keele-Essex-Dorset observations and
to within a percent for the N.\ Y.\ S.-Braeside datasets.  The
UKIRT data were reduced as described in Beekman et al
\shortcite{Beek97}.  The conversion to apparent magnitude was carried
out from observations of two UKIRT standards.  The lowest
signal-to-noise of any data point presented in this paper is 40, hence
we are confident that the data presented are of a high quality.  The
data were folded on the ephemeris of Beekman et al (1998);

\begin{equation}
BDJD = 2447965.884615 + 0.15820617 \times E.
\label{eq:ephem}
\end{equation}

Mid-eclipse is known to wander with respect to a linear ephemeris of up
to 300 s on timescales of 5000 cycles \cite{Wolf93}.

\section{The rise to outburst - `inside-out'}
\label{sec:phot}

\begin{figure*}
\begin{minipage}{170mm}
\vspace*{8cm}         

\includegraphics{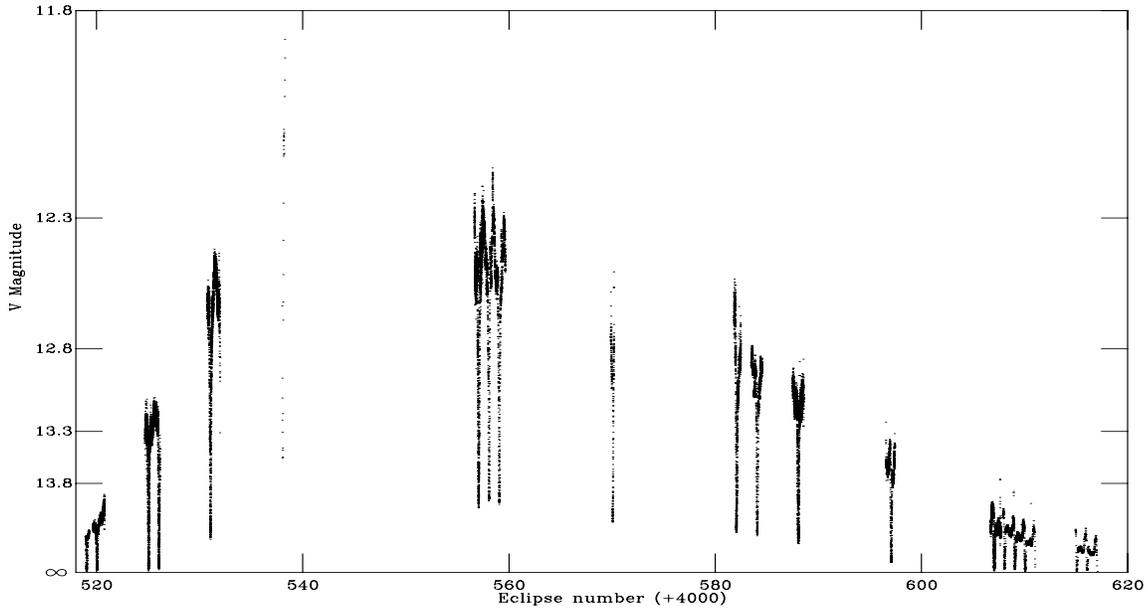}

\caption{The V band eclipses taken during the September - October 1997
outburst of IP Peg.}

\label{fig:all2}
\end{minipage}
\end{figure*}

\begin{figure*}
\begin{minipage}{170mm}
\vspace*{13cm}         

\includegraphics{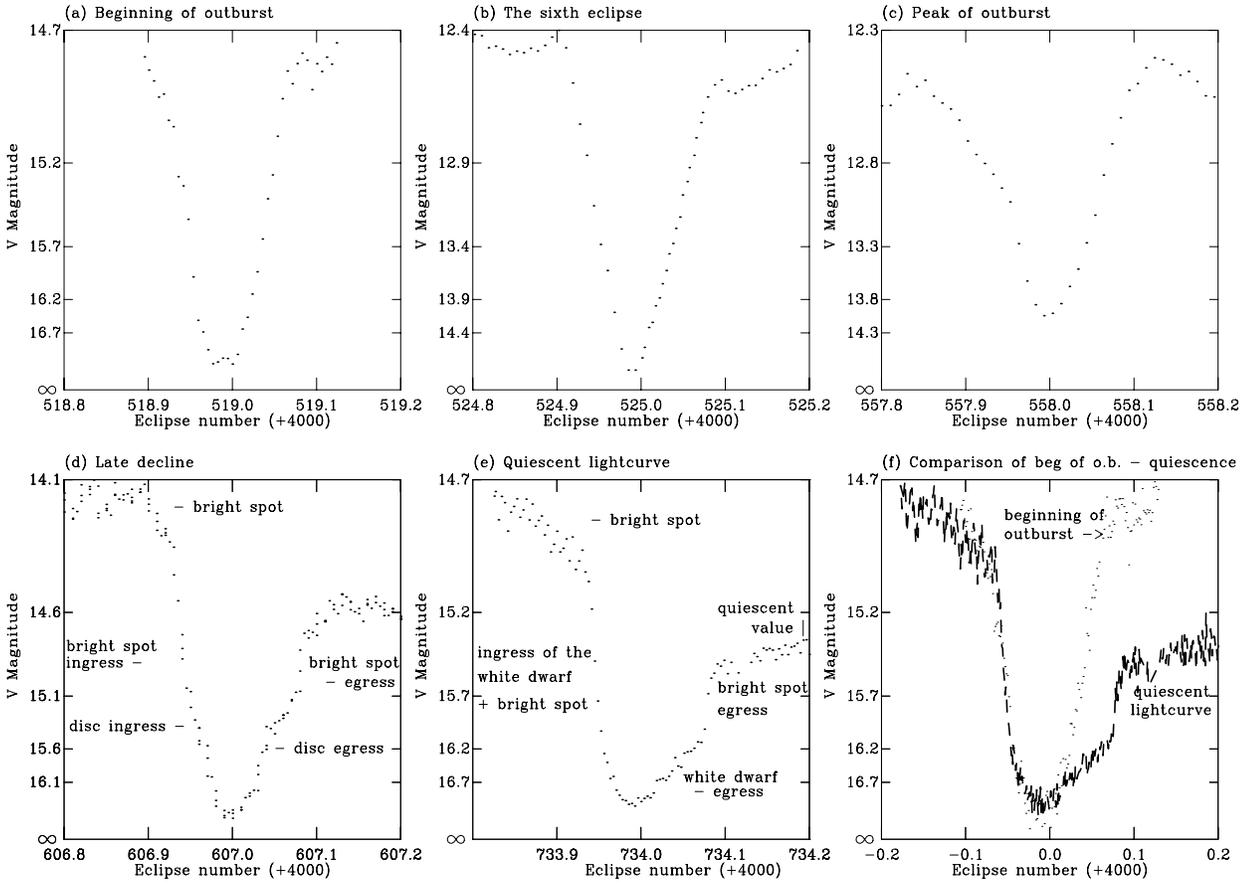}

\caption{Some of the eclipses binned up and expanded.  (a) is the first 
outburst eclipse, (b) is an eclipse from the second night.  (c) shows
a peak-of-outburst eclipse, (d) shows an eclipse from the late decline
phase and (e) shows a quiescent eclipse.  (f) compares the first
eclipse to the quiescent one and shows how little the system
brightness has changed at the very beginning of outburst.}

\label{fig:allcom}
\end{minipage}
\end{figure*}

\begin{figure}
\vspace*{5.3cm}         

\includegraphics{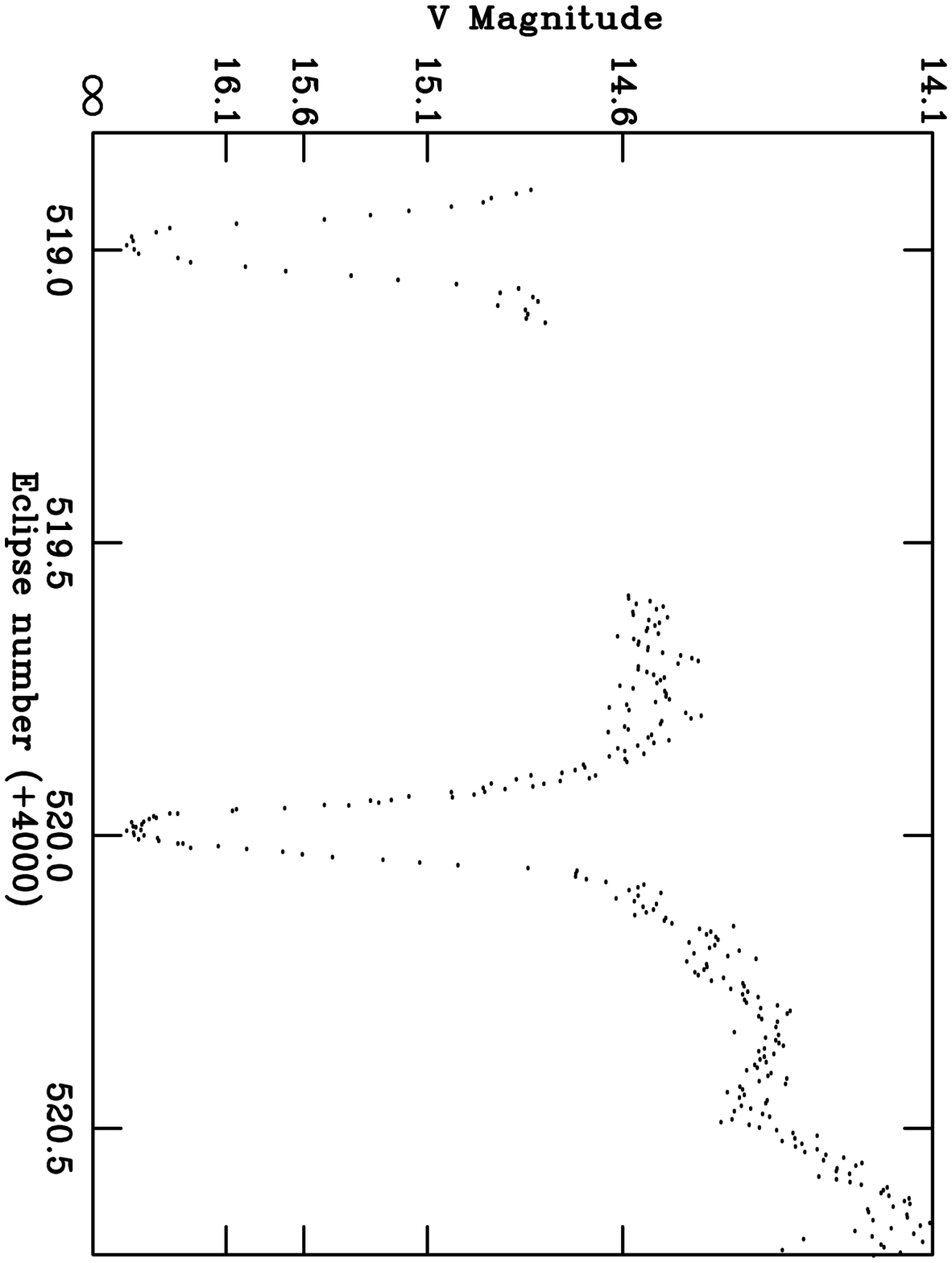}

\caption{Plot showing the first and second outburst eclipses (binned up).  The secondary eclipse is also visible at phase 0.5}

\label{fig:secec}
\end{figure}

The main result of this paper, that the outburst was `inside-out', can
be ascertained by consideration of Figs. \ref{fig:all2} and
\ref{fig:allcom}; although we shall show this in a more quantitative
way in Section~\ref{sec:discus}.  The primary evidence is the eclipse
depth, which decreases from 3.5 magnitudes in the first observations,
to less than 2 magnitudes at the peak-of-outburst.  At the onset,
there is little residual flux at mid eclipse, implying that the
secondary eclipses almost all the visible disc, with the small
residual flux probably coming from the secondary star \footnote{The
first and quiescent eclipses are very similar in magnitude, indicating
that the system was observed from the very beginning of outburst
(Fig.~\ref{fig:allcom}f).  Also the rapid rise in flux can be seen in
the out-of-eclipse region (Fig.~\ref{fig:secec}) in this initial
stage.}.  By the peak-of-outburst, the larger residual flux at
mid-eclipse suggests that the accretion disc has grown and become
visible around the limb of the secondary.

Further evidence for an `inside-out' outburst comes from the changes
in eclipse morphology.  The first eclipse has a flat bottom,
indicating that the object eclipsed is compact; see
Figs. \ref{fig:allcom}a and \ref{fig:secec}.  Even at the very start
of outburst, the bright spot is not very evident.  The early eclipses
are similar in width to the quiescent white dwarf eclipses, indicating
the visible disc is not large at this stage, compared to the
secondary.  However, the ingress and egress of the eclipses from the
beginning of outburst are not vertical (Fig.~\ref{fig:allcom}f),
suggesting that the central object is not entirely point like.  In
contrast, eclipses then become slightly `V-shaped' by the second night
(Fig.~\ref{fig:allcom}b).  By the peak-of-outburst they are V-shaped
but much shallower and wider at the top (see Fig.~\ref{fig:allcom}c).

\section{The decline from outburst}

We can follow the decline from outburst using eclipses 4607-10 and
4616-17.  In these, the bright spot reappears as a hump on the ingress
and emerging from eclipse on the egress (Fig.~\ref{fig:allcom}d).  The
eclipses become flat-bottomed, narrower and deeper again (see
Fig.~\ref{fig:allcom}d) and resume the form typical of an eclipsing
dwarf nova light curve (e.g Wood \& Crawford 1986), as IP Peg reaches
quiescence (Fig.~\ref{fig:allcom}e).  There are changes in the
gradient on many of the eclipse ingresses.  These probably mark the
beginnings and ends of the disc, bright spot and the white dwarf
ingress.  The position of the first change in gradient, attributed to
the disc, moves slowly towards mid-eclipse, indicating that the
feature causing this gradient change is becoming more centrally
condensed.  There is a persistent step at around phase 0.973 (see Fig.
\ref{fig:allcom}d).  This could be due to the white dwarf, as it is a
very compact object of static size.  There is, however, not always
evidence for the white dwarf reappearing on the egress at a similar
phase.  Also the `disc' egress is not always apparent, which makes the
disc size difficult to estimate.

\begin{figure}
\vspace*{5.3cm}         

\includegraphics{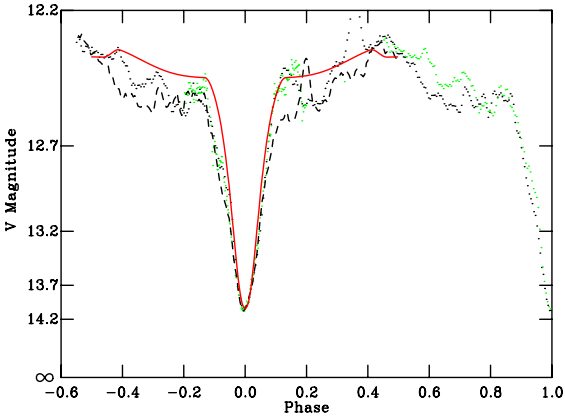}
\caption{Three consecutive eclipses taken from middle of outburst (binned up), solid line showing model fit (see Sect. \ref{sec:modelec}) to the average data.  Note the flickering in the out-of-eclipse regions.}
\label{fig:compare}
\end{figure}

\begin{figure}
\vspace*{5.3cm}         

\includegraphics{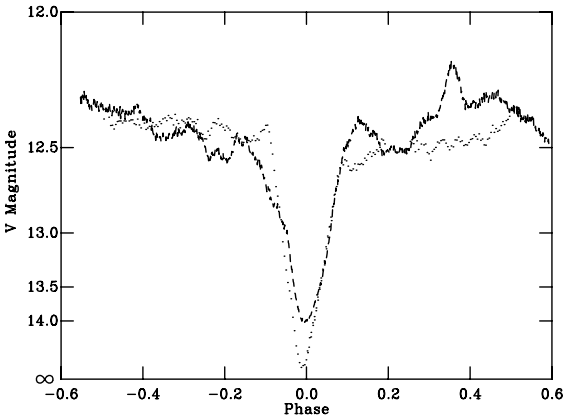}
\caption{Two eclipses compared (shifted for mid-eclipse to lie on phase=0), no's 4525 ($\times$ 1.08 mag), marked in dots, from the beginning of outburst and 4558, dotted line, from the middle.}
\label{fig:lessgood}
\end{figure}

\begin{figure}

\vspace*{5.3cm}         

\includegraphics{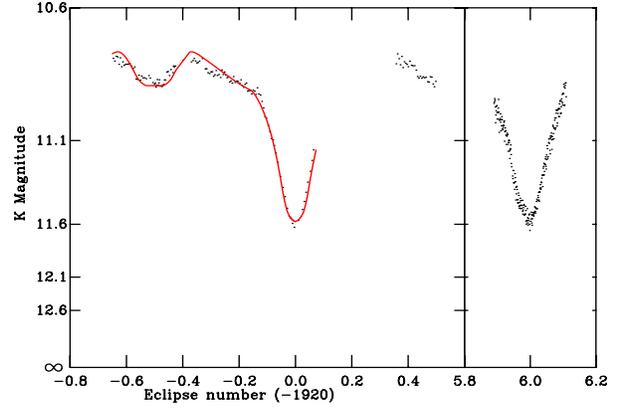}

\caption{Infrared lightcurves.  Some of the first nights data is folded on the orbital period and the data is binned up by a factor 5.  Model fit (see Section~\ref{sec:modelec}) is plotted over the top.}

\label{fig:ir}
\end{figure}

\section{Modelling the eclipses}
\label{sec:modelec}

\begin{table}
\caption{Fixed parameters used in the modelling code}
\label{tab:pars}
\begin{tabular}{lc}
Parameter & Value \\
\hline
Secondary pole temperature & 3100 K $^{a}$\\
Gravitational darkening coefficient & 0.08 \\
Orbital period of the system & 13669.0128288 s  \\
Mass of the primary & 0.9 M$_{\odot}$ \\
Distance & 150 pc \\
q & 0.49 \\
Inclination & 82$^{\circ}$ \\ 
\hline
\end{tabular}
$^{a}$revised Martin, Jones \& Smith, 1987 (only V band modelled)\\
\end{table}

To carry out further analysis of the outburst lightcurves, we
attempted to model them with the code described in Paper~I. This code
models the outburst lightcurves of dwarf novae.  We used it with the
parameters given in Table~2 fixed, to derive the parameters given in
Table~3.  One very important result found from modelling these
outburst eclipses, is that they are of a very different form to many
other dwarf novae outburst eclipses e.g. OY Car \cite{Rutt92} where
the eclipses in outburst are very symmetric from onset of the
outburst.  This unusual outcome is further explained in this section
and the following discussion section.  

Various q values were tried between 0.34-0.60, in steps of 0.01 (where
q=M$_2$/M$_1$).  The inclination and q value of the system are
related.  Assuming that the secondary fills its Roche lobe, the
duration of the eclipse of the white dwarf gives a constraint on the
relative sizes of the two component stars and the angle of inclination
(i) of the system to the line of sight \cite{Bail79}.  The inclination
can therefore be calculated from the q value and using a value of
$\Delta\phi$ (the smallest value of the white dwarf eclipse) from Wood
\& Crawford \shortcite{Wood86}.  For a q value of 0.39, i is required
to be 84$^{\circ}$, q=0.49 to be i=82$^{\circ}$ and q=0.58, i is
required to be 80$^{\circ}$.  0.39$\pm$0.05 \cite{Cata98} and 0.49
\cite{Wood86} gave comparable results.  That of Marsh (1998), q=0.58,
however, gave model fits with larger ${\chi^2}$ values than the lower
q values.  We used q=0.49 for all the modelling done in this paper.

\begin{figure}
\vspace*{5.3cm}   
\includegraphics{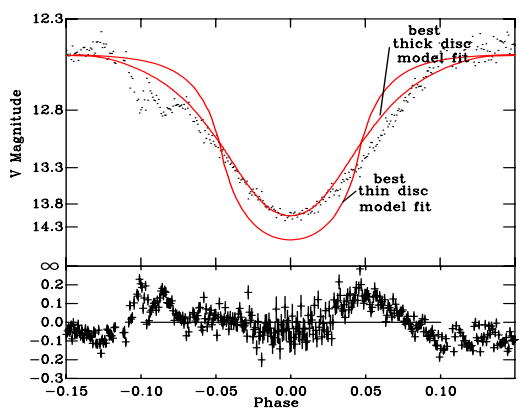}
\caption{Diagram showing a best fit (solid line) to the lightcurve 4557 (typical of many of the fits) and the residuals of the fit.  The best fit thin disc model is also plotted to demonstrate how poor a fit the thin disc model gives.}
\label{fig:modelfit}
\end{figure}

\begin{table*}
 \begin{minipage}{170mm}
 \begin{center}
 \caption{Results of modelling the disc radius, semi-opening angle of the disc and mass transfer rate through the disc of IP Peg throughout outburst. The final three columns have been calculated using the modelling code, for a mass ratio of q=0.49. Graph refers to the graphical method and model refers to modelling code method of determining disc radius. \hspace*{1.5cm} $^a$ see Section~\ref{sec:modelec}}
\label{tab:radii}
 \begin{tabular}{|lccccccccc|}
 \hline
 & Date & Eclipse & ${\bar{\phi}_d}$ & $\phi_d$ & r/R$_{L1}$  & r/R$_{L1}$  & r/R$_{L1}$  & Mass transfer & Semi-opening\\
 & (1997) & no.& & (max)&  (graph) & (max) &  (model) & through disc & angle ($^{\circ}$)\\
 &  & & & & &(graph) & & (x10$^{-9}$M$_{\odot}$/yr) & \\
Errors:& & & $\pm$0.001 & $\pm$0.001 & $\pm$0.002 & $\pm$0.002 & $\pm$5\% & $\pm$10\% & $\pm$20\% \\
\hline
 & Sep 21 & 4520 & 0.082 & 0.102 & 0.375 & 0.577  & 0.357 & 0.85 & 18\\
 & Sep 22 & 4525 & 0.102 & 0.112 & 0.563 & 0.650 & 0.481 & 2.75 & 16\\
 & Sep 23 & 4531 & 0.105 & 0.115 & 0.590 & 0.674 & 0.595 & 6.95 & 15\\
 & Sep 27 & 4557 & 0.114 & 0.131 & 0.662 & 0.793 & 0.64 & 8.00 & 15\\
 & Sep 28 & 4559 & 0.122 & 0.147 & 0.722 & 0.889 & - & -  & - \\
 & Oct 01 & 4582 & 0.084 & 0.092 & 0.400 & 0.472 & - & -  & - \\
 & Oct 02 & 4585 & 0.103 & 0.106 & 0.580 & 0.608 & 0.64 & 4.69 & 14\\
 & Oct 06 & 4608 & 0.093 & 0.098 & 0.482 & 0.528 $^a$ & - & -  & - \\
 \hline
 \end{tabular}
 \end{center}
 \end{minipage}
\end{table*}

The stream impact region (bright spot) manifests itself as a dark spot
during outburst \cite{Nayl87}.  This can be seen in
Fig.~\ref{fig:lessgood}, eclipse 4558 (typical of other eclipses),
where the ingress gradient is less steep initially.  As the cooler
region passes through the line of sight, it covers the central hot
part of the disc, thus less flux is observed at this point, hence the
eclipse begins earlier and is more gradual, giving an asymmetric
profile.

The eclipses could be fit in depth and breadth, but on both the
ingress and the egress, the fit deviated from the data systematically,
where the eclipse shape was irregular, in part due to the bright/dark
spot.  Owing to the poor fit, the $\chi$$^2$ values given by the model
are much larger than 1 and so appear inappropriate for testing the
validity of the fit (Fig.~\ref{fig:modelfit}).  However, they are
still useful in pointing us towards the better fits of the model to
the data.  The disc radius and opening angle have large errors, as
there is a `trade-off' between the two quantities when fitting the
model.  The data can be fitted with a smaller radius and a thicker
disc, or a larger but thinner disc.  However, by fitting the data with
two models that fit just below and just above the eclipse data points,
we estimated the errors given in Table~3.  The results show a general
trend of the disc radius and mass transfer rates throughout the
outburst.

To obtain a more consistent set of results regarding the change in
radius of the luminous part of the disc, the first and last contact
points were measured for each eclipse. The contact points were
difficult to discern because of flickering, so the gradient between
successive data points of the eclipse was plotted against phase, from
which the contact points could be more easily distinguished. Hence the
results can only be considered to be accurate to $\pm$0.001 in phase.
The eclipse widths, in phase, are given in Table 3. ${\bar{\phi}_d}$
is half the difference in phase from the first to the last contact
point.  $\phi_d$(max) is the larger of the two phases between the
first contact point and phase zero, or between phase zero and the last
contact point.  Table 3 shows that the width of the eclipse increases
over the rise to outburst and decreases as the system declines from
outburst.  The asymmetry of the disc can also be seen.  r/R$_{L1}$ values are
generally more extreme than the model values, as the model assumes a
symmetric disc.  However, both sets of results follow the same trend.

The radius of the disc can be estimated from the first and last
contact points of the eclipse in the following way. The modelling code
was used to create a model for discs of 0.2R$_{L1}$, 0.4R$_{L1}$,
0.6R$_{L1}$ etc using a thin disc.  The phase between the first and
last contact points derived from the model lightcurve was plotted
against the radius of the thin disc.  The radius of the disc for each
phase value measured was then read from the graph. The results of this
are given in Table 3.  This was the best method for deriving disc
radius that we tested as it did not rely on an interdependency between
the disc radius and opening angle and shows the assymmetry of the
disc, which the model is unable to do.  The same trend in disc size
variation throughout the outburst was determined, as had been found
from the modelling \footnote{It should be noted, that if, at the end
of outburst, only the final part of the ingress and the first part of
the egress are associated with the disc, as described in
Section~\ref{sec:phot}, the disc size at the end-of-outburst is
considerably smaller than given in the table, approaching the size of
the white dwarf.}.

\section{Discussion}
\label{sec:discus}

\subsection{Comparison with the $\alpha$ disc theory}

The $\alpha$ disc theory asserts that long rise times, i.e. long
compared with other outbursts of IP Peg, are indicative of inside-out
outbursts.  This is due to the rise time being related to the time
taken by the heating front to traverse the disc.  An outward moving
front moves more slowly than an inward moving one, hence longer rise
times.  The outbursts of IP Peg exhibit a range of rise times.  From
the British Astronomical Association Variable Star Section (BAAVSS)
database, we have estimated rise times for 13 IP Peg outbursts between
November 1991 and January 1998.  The rise time estimates fall into two
categories of either less than two days (short rise time) or three
days and more (long rise time).  Since this outburst has a long rise
time of about three days, it is consistent with the theoretical
prediction that long rise times are associated with inside-out
outbursts.  The manner in which the disc is small at the beginning of
outburst, compared to the tidal radius and then approaches the tidal
radius value during the outburst, then slowly falling to the initial
radius is consistent with previous analysis of IP Peg outbursts,
e.g. Wolf et al (1993), although previously analysis has used no rise
to outburst data, only the fact that the disc was small prior to
outburst.  This is in sharp contrast to a system showing `outside-in'
outbursts, e.g. HT Cas (Paper~I), where the luminous part of the disc
is initially large and then decreases in radius through the outburst.

\subsection{Disc edge structure}
\label{sec:des}

The tidal radius of the system can be calculated by first finding
R$_{\rm{L1}}$/a numerically, then using the equation of Paczynski
\shortcite{Pacz77}.   For a q value of 0.39, the tidal radius is
0.727 R$_{L1}$ and for a q value of 0.49, the tidal radius is
0.705R$_{L1}$.  From modelling, the luminous disc radius is well below
the tidal radius at the beginning of outburst and so we expect no
tidal interaction between the secondary star and the disc.  

At the peak-of-outburst, model values of the size of the accretion
disc are within 0.05 R$_{L1}$ of the tidal radius.  However, this is
the average size of the disc.  The disc is highly asymmetric (see
average and peak values of the first and last contact points in Table
4).  By the peak-of-outburst at least some of the disc has reached the
tidal radius and is presumably disrupted by the secondary star.  This
point in the outburst coincides with the maximum flickering observed
in the light curves outside of eclipse, see Figs.~\ref{fig:compare}
and \ref{fig:lessgood} (the latter of which shows an early eclipse
with little flickering (no. 4525) compared to a later one (no. 4558)
from the middle of outburst.  This could show that the disc edge has
become disrupted at this point, as the disrupted edge has thicker and
thinner regions.  When a thicker part of the edge of the disc passes
our line of sight, less of the hot central part of the disc is visible
and so we receive less light (a dip in the light curve) and conversely
for a peak in the lightcurve.  Another possible explanation is that
the flickering is due to matter being accreted onto the white dwarf in
a clumpy fashion e.g. \cite{Gott91}.  The flickering in the
non-eclipsed part of the light curve diminishes during the decline
phase.

\subsection{The secondary star}
\label{sec:secstar}

The secondary star is evident in both the V and K band data. The
secondary eclipse is visible at phase 0.5 at the beginning of outburst
in the V band (depth 0.05 mag.), when the disc has not
brightenend sufficiently to outshine the secondary (see
Fig.~\ref{fig:secec}).  The eclipse of the secondary star is also
visible in the K band peak-of-outburst lightcurves,
(Fig.~\ref{fig:ir}), where it lasts 0.2 in phase and is 0.1
magnitudes deep.  This is superimposed on a broad hump, centred on
phase 0.5, which covers more than half the orbital cycle.  We believe
that this hump is due to the irradiated face of the secondary star,
rotating into view.  

\begin{figure}

\vspace*{3.3cm}         

\includegraphics{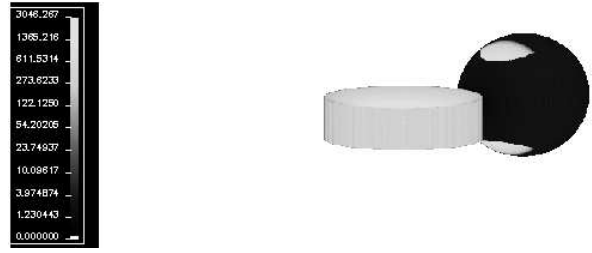}

\caption{Irradiation of the secondary at the peak-of-outburst, phase
0.4.  The greyscale is flux (erg cm$^{-2}$ s$^{-1}$ \AA$^{-1}$
sterad$^{-1}$) at the surface towards the observer.}

\label{fig:irrad}
\end{figure}

To test this hypothesis, we modelled the effect of irradiation of the
secondary star in both the K and V bands.  We placed an irradiating
source at the centre of the disc and modelled the luminosity required
to reproduce the lightcurves in the K band, using the modelling code
as in Section~\ref{sec:modelec}.  The thickness of the disc caused
some shadowing of the secondary, but the upper and lower parts of the
inner face were irradiated (Somers, Mukai \& Naylor, 1996 and
Fig.~\ref{fig:irrad}).  From this modelling we found that a luminosity
of $\simeq$10L$_{\odot}$ must be irradiating the secondary.  Such high
luminosities are not generally known to emanate from white dwarfs (see
Sion 1995); so it is highly unlikely to be coming from the white
dwarf.  From the mass transfer rates given in Table 3, accretion
luminosities from the boundary layer, similar to the observed
luminosity are produced.  Thus the boundary layer is probably the
source of irradiation.  

To show that the irradiation model is consistent with the V band data,
we fitted the averaged V band lightcurve from the peak-of-outburst,
see Fig.~\ref{fig:compare}.  This was also well fitted with a
luminosity of $\simeq$10L$_{\odot}$.  In the V band model, the broad
hump was also evident, superimposed over a small eclipse centered at
phase 0.5.  This eclipse of the secondary star is not evident in the
peak-of-outburst data, as the flickering of 0.2 magnitudes is larger
than the secondary eclipse.  Thus we see evidence of the secondary
star, both from its eclipse in the beginning of outburst data and the
modulation due to its irradiated face.  The irradiating source,
probably the boundary layer, has a luminosity of
$\simeq$10L$_{\odot}$.

The model of the out-of-eclipse V band lightcurve is almost horizontal
as the disc comes out of eclipse (Fig.~\ref{fig:compare}).  It then
shows a very slow rise until phase $\sim$0.2, when it increases much
more quickly.  This is because very little of the irradiated region of
the secondary is visible until phase 0.12, so there is little increase
in flux.  At phase 0.16, both the upper and lower part of the
irradiated face become visible, see Fig.  \ref{fig:irrad}.  Of this,
the hottest region becomes visible at phase $\sim$0.22. Then the upper
and lower irradiated regions begin to contribute a significant amount
of flux in the V band - hence the gradient of the lightcurve increases
more rapidly.  Although the secondary eclipse begins at  phase
$\sim$0.38, it is only evident at phase $\sim$0.41 in the V band, as this is
when the irradiated regions of the secondary begin to be eclipsed.
From modelling, the irradiated region on the secondary is as hot as
all but the very central parts of the disc, hence its contribution to
the lightcurve (Fig.~\ref{fig:irrad}).

\subsection{Spiral arms}
\label{sec:sparms}

Although we have no spectroscopy of IP Peg, from which in previous
analysis spiral shocks have been identified, we analysed the light
curves as suggested by Bunk, Livio \& Verbunt \shortcite{Bunk90}.
They suggest that light curves for optical data, where the spiral arms
are in the orientation suggested by Steeghs, Harlaftis \& Horne
(1997), should show asymmetric eclipse profiles, with a steep ingress
gradient and a shallow egress gradient and a mid-eclipse region
between ingress and egress which is relatively flat, albeit with a small
negative gradient.  We found similarly formed ingress and egress
gradients in this dataset, but without the flat bottom.  However, this
profile is not only observed for the eclipses at the peak-of-outburst,
where the disc is thought to have reached the tidal radius, but in all
other eclipses \footnote{Similarly formed eclipses from other
eclipsing Cataclysmic Variables can also be observed, where the
eclipse shape is attributed to the bright spot, as pointed out by the
referee.}  .  The steep ingress and shallower egress are also typical
of arms at orientations other than those postulated by Steeghs,
Harlaftis \& Horne (1997).  However, it must also be remembered that
these models \cite{Bunk90} are for a thin disc and we have shown that
the disc at peak-of-outburst is flared, to $\simeq$ 15$^{\circ}$.  In
addition, further out-of-eclipse evidence should be observable.  If
the spiral arms exist and are orientated as described by Steeghs,
Harlaftis \& Horne (1997) centered around phases 0.25 and 0.75, then
evidence of their presence should be observed at these points on the
light curve.  We folded five light curves (binned together to remove
irregular structure in the out-of-eclipse region) from the
peak-of-outburst (when the system has reached the tidal radius) and 5
lightcurves where the disc is much less than the tidal radius.  We
examined the resultant light curves, but found no evidence for
features at these phases that could not be explained by the irradiated
secondary.

What our data show is that from simple photometry only, the existence
of spiral arms cannot be proved, although the data are not
inconsistent with their existence.  However, little work has been done
to show how models of discs that rely only on spiral arms as their
mode of transporting angular momentum outwards can go into outburst.
It has not yet been proved that spiral arms would provide a viable
mechanism to transport all of the angular momentum away in cooler
and/or thick discs.  In all models that have been produced, the method
for transporting angular momentum away uses unrealistically hot discs
(large speed of sound) and the cooling effect has been neglected
\cite{Livi91}.  Further supporting evidence is therefore required from
both observation and theory in order to prove their existence.  In
contrast, our data can be entirely explained without the need for
spiral arms, using the `inside-out' $\alpha$ disc instability model.
This could question the need for spiral arms in the discs of
dwarf novae.

\section{Conclusions}

We have shown that the outburst undergone by IP Peg in September 1997
was of the `inside-out' variety.  As part of this evidence, we show
that the visible disc size grows in radius over the rise to outburst
and then falls over the decline.  From our modelling, due to large
errors we cannot determine a trend in opening angle over the outburst,
but the visible radius increases from about 0.5R$_{L1}$ to
0.7R$_{L1}$.  The mass transfer rate also follows the same trend.
After increasing from about 2.2x10$^{-10}$M$_{\odot}$yr$^{-1}$ in
quiescence \cite{Mars88a}, to 8.5x10$^{-10}$M$_{\odot}$yr$^{-1}$ at
the beginning of the outburst, it then reaches at least
8.0x10$^{-9}$M$_{\odot}$yr$^{-1}$ at the peak of outburst.  These
values should be viewed with caution, as the errors are large, though
they do show the trend in the data.  The disc reaches the tidal radius
around the peak-of-outburst and is highly asymmetric.  We also show
results that are consistent with the theoretical prediction that long
rise times are features of `inside-out' outbursts.  We have shown
evidence for the irradiation of the secondary star by the white dwarf
and/or more likely the boundary layer.  The irradiating luminosity is
of the order of 10L$_{\odot}$.  We have, however, no evidence to
support the idea that the disc contains spiral arms and we can fully
explain our lightcurves using only the $\alpha$ disc instability
model.

\section{Acknowledgements}

We acknowledge the BAAVSS for use of their archive,  TN was in
recipient of a PPARC advanced fellowship during some of this work and NAW
is supported by a PPARC studentship. We are also grateful to
G. Beekman for helping us obtain data with UKIRT and for many
informative discussions.  The United Kingdom Infrared Telescope is
operated by the Joint Astronomy Centre on behalf of the U.K. Particle
Physics and Astronomy Research Council.  We also acknowledge the
amateur astronomers, who maintain the Keele telescope.  The data
reduction was carried out on the Keele STARLINK node using ARK
software.

\end{document}